# Magnetoelectric ordering of BiFeO$_3$ from the perspective of crystal chemistry

L. M. Volkova, D. V. Marinin

Institute of Chemistry, Far Eastern Branch of Russian Academy of Science, 690022 Vladivostok, Russia

**Abstract**

In this paper we examine the role of crystal chemistry factors in creating conditions for formation of magnetoelectric ordering in BiFeO$_3$. It is generally accepted that the main reason of the ferroelectric distortion in BiFeO$_3$ is concerned with a stereochemical activity of the Bi lone pair. However, the lone pair is stereochemically active in the paraelectric orthorhombic ß-phase as well. We demonstrate that a crucial role in emerging of phase transitions of the metal-insulator, paraelectric-ferroelectric and magnetic disorder-order types belongs to the change of the degree of the lone pair stereochemical activity – its consecutive increase with the temperature decrease. Using the structural data, we calculated the sign and strength of magnetic couplings in BiFeO$_3$ in the range from 945º down to 25º C and found the couplings, which undergo the antiferromagnetic→ferromagnetic transition with the temperature decrease and give rise to the antiferromagnetic ordering and its delay in regard to temperature, as compared to the ferroelectric ordering. We discuss the reasons of emerging of the spatially modulated spin structure and its suppression by doping with La$^{3+}$.

## 1 Introduction

The role of crystal chemistry factors in emerging of the relation between the electric and magnetic orderings in multiferroics is not yet absolutely clear. On the other hand, determination of the crystal chemistry conditions under which the above takes place is important in search of the compounds – potential multiferroics characterized by these very conditions.

In view of the above, the crystal chemistry study of the BiFeO$_3$ multiferroic is of special interest, since it has extremely high temperatures of ferroelectric ($T_C$ ~ 810-830º C) [1, 2] and antiferromagnetic ordering ($T_N$ ~ 350-370ºC) [3, 4], which is very important for technological applications. At the same time, bismuth ferrite has a substantial disadvantage. It was stated [5-7] that the non-collinear magnetic structure of bismuth ferrite comprising a modulated spin structure with a large period prevents from emerging of the linear magnetoelectric effect and spontaneous magnetization. One of the ways of suppressing the modulated spin structure is the application of a strong magnetic field [8-10]. However, in such a case the values of field strengths are so high (~200 kOe) that a practical application of the "non-evident magnetoelectric" BiFeO$_3$ has some difficulties. Another ways of suppressing the modulated spin structure consist in epitaxial constraint and/or substitution of bismuth ions by rare earth elements [11-17]. The presence of lanthanum additives results in the decrease of the field of transition from the spatially modulated structure to the uniform antiferromagnetic one and, therefore, in elimination of one of the main disadvantages of the BiFeO$_3$ multiferroic.

The experimental studies of BiFeO$_3$ [18-20] established that the temperature increase produced consecutive structural phase transitions from the ferroelectric α-phase (rhombohedral space group $R3c$ with $a$ = 5.6 Å, $c$ = 13.9 Å and γ=120) [5, 21-27] to two paraelectric phases. First, the process leads to the orthorhombic ß-phase [19] (space group $Pbnm$ with $a$ = 5.613 Å, $b$ = 5.647 Å, and $c$ = 7.971 Å) between 820º C and 830º C and then, in the range 925-933º C, to the cubic γ-phase [18] (space group $Pm\bar{3}m$ with $a$ = 3.992 Å), which decomposes above 960º C. The transition to the cubic phase is accompanied by the insulator-metal transition [18]. Nevertheless, there is still no agreement among the researchers on the transition temperature and symmetry of the paraelectric γ-phase. The crystal structure of the γ-phase has been identified as rhombohedral $R\bar{3}c$ in [28], tetragonal $I4/mcm$ in [29], cubic $Pm\bar{3}m$ in [30], monoclinic $P2_1/m$ or $C2/m$ in [31], and, recently, as orthorhombic $Pbnm$ in [32]. The change of structure, loss of magnetic order and metallization of BiFeO$_3$ are also observed under pressure of 45-55 GPa at room temperature [33-36].

To sum up, it has been established that the increase of any of the parameters (temperature and pressure) produced the same effects: the magnetic ordering is eliminated and the ferroelectric-paraelectric and, further, the insulator-metal phase transitions take place. These transitions are related to the structural changes. It was concluded on the basis of the performed studies [37-44] that the main reason of the ferroelectric distortion in BiFeO$_3$, just like in PbTiO$_3$ and BiMnO$_3$, consists in the stereochemical activity of the so-called *lone pairs* – two valence *s*-electrons of Bi$^{3+}$ and Pb$^{2+}$ cations. The polarization analysis performed in [42] demonstrates that partial contributions to polarization from the Fe and O atoms almost cancel each other and the net polarization present in BiFeO$_3$ originates mainly (>98%) from Bi atoms. This conclusion is not supported in [36, 45], where the authors indicate the high-spin–low-spin crossover in the electronic *d* shell of 3*d* transition metal ion Fe$^{3+}$ with $d^5$ configuration as a primary source of all phase transitions under high pressures. However, in [46] the latter statement is disproved and it is again believed that both the temperature-driven and the pressure-driven insulator-metal transitions are controlled by the crystallographic change with the key structural parameter being the Fe-O-Fe bond angle and not by the electron-electron repulsion.

In the present work we have performed the following studies:
- detailed analysis of the structural aspect of ferroelectric-paraelectric and insulator-metal phase transitions in BiFeO$_3$ and demonstration of the above transitions relation to consecutive change of the Bi lone electron pair behavior, under the temperature effect, from high stereochemical activity to a reduced one and, finally, to a stereochemically inert state;
- calculation, on the basis of the structural data at temperatures from 25º up to 945º C, of the sign and strength of magnetic interactions in BiFeO$_3$ and demonstration of the relation between the change of magnetic properties and the change of geometric positions of intermediate ions in the local space between the iron ions for the nearest- and next-nearest-neighbour magnetic interactions under the temperature effect;
- determination of crystal chemistry factors facilitating the emerging of electric and magnetic ordering in BiFeO$_3$.

## 2 Method

To determine the characteristics of magnetic interactions (type of the magnetic moments ordering and strength of magnetic coupling) in the multiferroic BiFeO$_3$ and search for novel multiferroics, we used the previously developed phenomenological method [47, 48], which we named the *"crystal chemistry method"*, and the program "MagInter" developed on its basis. The method enables one to determine the sign (type) and strength of magnetic couplings on the basis of structural data. According to this method, a coupling between magnetic ions $M_i$ и $M_j$ emerges in the moment of crossing the boundary between them by an intermediate ion $A_n$ with the overlapping value of ~0.1 Å. The area of the limited space (local space) between the ions $M_i$ and $M_j$ along the bond line is defined as a cylinder, whose radius is equal to these ions radii. The strength of magnetic couplings and the type of magnetic moments ordering in insulators is determined mainly by the geometrical position and the size of intermediate $A_n$ ions in the local space between two magnetic ions $M_i$ and $M_j$. The positions of intermediate ions $A_n$ in the local space are determined by the distance $h(A_n)$ from the center of the ion $A_n$ up to the bond line $M_i$-$M_j$ and the degree of the ion displacement to one of the magnetic ions expressed as a ratio ($l_n'/l_n$) of the lengths $l_n$ and $l_n'$ ($l_n \leq l_n'$; $l_n' = d(M_i - M_j) - l_n$) produced by the bond line $M_i$-$M_j$ division by a perpendicular made from the ion center. The intermediate ions included into the local space between magnetic ions $M_i$ and $M_j$ would, depending on their positions, orient the magnetic moments of these ions and contribute $j_n$ into emerging of antiferromagnetic (AFM) or ferromagnetic (FM) components of the magnetic interaction. The sign and strength of the magnetic coupling $J_{ij}$ is determined by the sum of the above contributions:

$$J_{ij} = \sum_n j_n \qquad (1)$$

The value $J_{ij}$ is expressed in units of Å$^{-1}$. The comparison of our data with that of other methods shows that the scaling factor for translating value angstrom$^{-1}$ into meV in oxides Fe$^{3+}$ is 38.5 (Table 1). If $J_{ij} < 0$, the type of $M_i$ and $M_j$ ions magnetic ordering is AFM and, in opposite, if $J_{ij} > 0$, the ordering type is FM.

**Table 1** An estimate of $J_1$ magnetic couplings in oxides Fe$^{3+}$ by crystal chemical method (I) and experimental and quantum-chemical methods (II)

| Compound | Space group, lattice parameters | d(Fe-Fe) (Å) | $J_1$ (Å$^{-1}$) (I) (AFM<0) | 38.5×J(Å$^{-1}$) | $J_1$ (meV) (II) (AFM>0) |
|---|---|---|---|---|---|
| KFe$_3$(SO$_4$)$_2$(OH)$_6$ [49] | $R\bar{3}m$ (N166) $a$ = 7.311 Å, $c$ = 17.175 Å, $\gamma$ = 120º | 3.656 | -0.0826 -0.0934**b** | -3.18 -3.60 | 3.225 [50], 3.900 [51, 52] |
| Cu$_2$Fe$_2$Ge$_4$O$_{13}$ [53] | $P2_1/m$ (N11) $a$ = 12.088 Å, $b$ = 8.502 Å, $c$ = 4.870 Å, $\beta$ = 96.17 | 3.208 | -0.0443 | -1.71 | 1.60(2) [54], 1.7 [54] |
| SrFeO$_2$ [55] | $P4/mmm$ (N123) $a$ = 3.991 Å, $b$ = 3.991 Å, $c$ = 3.475 Å, | 3.991 | -0.1758 | -6.77 | 7.04 [56] |
| γ-BiFeO$_3$ [18] | $Pm\bar{3}m$ (N221) $a$ = 3.992 Å | 3.992 | -0.1757 | -6.76 | 6.3 [57], 6.54 [58], 7.1 [57], 7.4 [58, 40] |

The value of the contribution to the AFM or FM coupling components is maximal, if the intermediate ion position is in the central one-third of the local space between magnetic ions. For the maximal contribution into the AFM-component of the coupling, the intermediate ion must be at the closest distance from the axis, while in the case of the FM-component, in opposite, from the surface of the cylinder limiting the space area between the magnetic ions. The distance between the magnetic ions $M_i$ and $M_j$ affects only the value of contribution, but does not determine its sign. We have assumed in our calculations the coupling strength to be reverse-proportional to the square of the distance between the magnetic ions $M_i$ and $M_j$. However, the dependence of the coupling strength on the above distance is more complicated. With increasing the distance, the coupling strength decrease occurs at a higher rate. Juxtaposition of the data obtained using our method with experimental results of studies of the known magnetic compounds showed that the coupling strength was reverse-proportional to the distance square at the d($M_i$-$M_j$) distance increase up to ~8 Å, while during further increase of the distance the coupling strength must be reverse-proportional not to the square, but to the cube of the distance. However, the available literature does not contain sufficient reliable data to take the above effect into account in our method. As a result, the strength of couplings between ions located at long distances might be artificially elevated.

One of the main disadvantages of the mentioned method is concerned with using the parameters considered as 'effective', but not as strictly defined constants. These parameters include atomic and ionic radii, which, according to quantum chemistry studies, do not have definite physical meaning. Nevertheless, the 'sizes' of atoms and ions are successfully used for prediction and explanation of structural effects at substitutions and correlate with many physical properties of compounds.

The method is sensitive to insignificant changes in the local space of magnetic ions and enables one to find intermediate ions localized in critical positions, deviations from which would result in the change of the magnetic coupling strength or spin reorientation (AFM-FM transition).

In the BiFeO$_3$ samples two critical positions – 'a' and 'd' – can be observed [47, 48]. In the 'a' position ($h_c(A_n) = r_M + r_{A_n} - 0.1$) the ion $A_n$ enters the local space by less than ~0.1 Å and does not initiate the emerging of magnetic interaction ($j_{A_n} = 0$). However, at slight decrease of the distance $h(A_n)$ from the $A_n$ ion center to the bond line Fe$_i$-Fe$_j$ (the $A_n$ ion displacement inside this area) there emerges a substantial contribution of this ion to the coupling FM component. It appears difficult to determine the value of the critical distance $h_c(A_n)$ with high accuracy from the structural data, since the values of as ions radii as atom coordinates contain some errors. The critical distances from the ion centers O, Bi and Fe to the bond line Fe$_i$-Fe$_j$ are equal to ~1.95 Å, ~1.72 Å and ~1.19 Å, respectively. The position 'd' emerges in the case when several intermediate ions $A_n$ are located between the magnetic ions $M_i$ and $M_j$ and the sums of these ions contributions $j_n^s$ to AFM and FM components are approximately equal making the magnetic coupling weak and unstable. Small displacement of any of the intermediate ions might result in the coupling complete disappearance or AFM-FM transition.

The sign and strength of magnetic couplings in BiFeO$_3$ was calculated in the temperature range from 25º up to 945º C. We took the crystallographic parameters and atom coordinates in γ-BiFeO$_3$ [18, 32], β-BiFeO$_3$ [19], α-BiFeO$_3$ [21] and Bi$_{0.93}$La$_{0.07}$FeO$_3$ [24], as well as the ionic radii of Fe$^{3+}$, Bi$^{3+}$ and O$^{2-}$ ($r_{Fe^{3+}}$ = 0.645 Å, $r_{Bi^{3+}}$ = 1.17 Å (for γ- and β-BiFeO$_3$), $r_{Bi^{3+}}$ = 1.103 Å (for α-BiFeO$_3$), $r_{O^{2-}}$ = 1.40 Å) determined in [59], as initial data for calculations. The method high sensitivity to any slight change of atom coordinates comprises simultaneously its advantage and disadvantage. That is why the errors and mistakes in the crystal structure determination will be also inherent to the calculation results obtained using this method. The crystallographic data and calculated characteristics of magnetic interactions BiFeO$_3$ are presented in Tables 2 and 3.

## 3 Changes in the lone pair behavior in BiFeO$_3$ under the temperature effect

The effects related to the lone pair of electrons belonging to a cation have not yet found an unambiguous explanation. For the first time the stereochemical effect of the lone pair was represented as a result of the valence shell electron pair repulsion (VSEPR) [60]. On the basis of the mentioned approach, Gillespie and Nyholm [61, 62] developed a set of rules that enabled one to interpret experimental data and predict the structure of compounds of incomplete-valency post-transition elements. Later Orgel [63] put forward the hypothesis that the structural distortion resulted from the mixing of the metal cation *s*- and *p*-orbitals. However, Waghmare et al. [41], by using the first principles of the density functional theory for studies of the IV-VI chalcogenide series, determined the detailed chemical composition of lone pairs and established the factors that caused lone pairs to favor high- or low-symmetry environments. They concluded on the basis of the above that the traditional picture of cation *s-p* mixing causing localization of the lone pair lobe was incomplete, and instead the *p* states on the anion also play an important role. In addition, these compounds reveal a delicate balance between two competing instabilities – structural distortion and tendency to metallicity. The same conclusion was made by Watson et al. [64] in performing the *ab initio* calculation of the origin of the distortion of α-PbO. Analysis of the partial density of states reveals mixing of the Pb 6*s* with the oxygen 2*p* electronic states indicating that the classical theory of hybridization of the lead 6*s* and 6*p* orbitals is incorrect and that the lone pair is the result of the lead-oxygen interaction. These two very studies [41, 64] formed a basis for the explanation of the emerging of ferroelectricity in Bi and Pb perovskites given by Hill [37], Neaton [65], and Ravindran [42]. They demonstrate that the ferroelectricity is originating from the distortion of Bi(Pb)-O coordination environment as a result of the stereochemical activity of the lone pair on Bi(Pb). According to their calculations, the lone pair includes not only Bi(Pb) 6*s* and 6*p* states, but also has some contribution from the 2*p* states on the oxygen ligands. Moreover, the key role in the ferroelectricity stabilization is attributed to covalent bonding between Bi(Pb) cations and oxygen anions. Atanasov and Reinen [66, 67] explained the lone pair stereochemical activity with using the vibronic coupling model.

Besides, there are other opinions on the formation and behavior of the lone pair. Lefebvre and co-workers [68, 69] showed on the basis of studying the electronic structure of antimony and tin chalcogenides that this pair did not take part in the bonding but tended to expand as far as possible, distorting the anionic arrangement around cations, while anions had also a great influence on the lone pair behavior. According to Khomskii [43], the main instability leading to ferroelectricity in perovskites BiMnO$_3$ and BiFeO$_3$ is driven by the Bi$^{3+}$ ions, whose lone pairs in the systems under study do not participate in bonds. The latter produces high polarizability of the Bi$^{3+}$ ions, which strongly enhance the instability towards ferroelectricity. The particular orientation of these lone pairs, or dangling bonds, may create local dipoles, which finally can order in a ferroelectric or antiferroelectric fashion.

The crystal chemistry does now allow determining the electronic structure of a compound; it just handles its result – the crystal structure. Independently of the way of the lone pair formation, it is clear that its role in formation of the compound structure and properties is substantial.

The changes in the lone pair behavior under the temperature effect can be most clearly demonstrated within the scopes of the geometrical model of the valence shell electron pair repulsion (VSEPR) [61] by taking this model's grounding postulate on the competition for the position near the atom skeleton between the coupling and lone electron pairs while rejecting the point on the lone pair as a rigid sphere. We showed earlier [70, 71] in our studies of the stereochemical role of the lone pair in the compounds of Sb$^{3+}$ and Bi$^{3+}$ that the lone pair, unlike the coupling pairs localized between two atom skeletons, had three positioning options: first, it could occupy a localized site near the atom skeleton (as ligand); second, it could be uniformly distributed (delocalized) around the skeleton; third, it could occupy an intermediate position between the first two. In the first case the lone pair is considered stereochemically active and its structural effect is reflected in the presence of two spheres: ligands are located at short distances on one side of the cation (sphere I); the electron cloud of the lone pair surrounded by additional ligands at long distances is located on the other side of the cation (sphere II). In the second case the lone pair is considered stereochemically inert, since its coordination geometry does not distort, while the structural effect consists in a uniform repulsion of coupling orbitals resulting in increase of all M-X distances as compared to short distances found in the compounds of these elements having a stereochemically active lone pair. We showed in [71] that the increase of pressure or temperature or the number of ligands and/or decrease of their electronegativity resulted in the decrease of the lone pair stereochemical activity. The latter is expressed in the decrease or complete elimination of the cations coordination surrounding distortion with the lone pair and produces structural phase transitions and changing of the respective physical properties. The ability of the lone pair to change its shape and position was shown through visualizing the electronic structure changes in $Pb_2MgWO_6$ [72] at the phase transition from a cubic paraelectric phase to an orthorhombic antiferroelectric phase below 310K. The above changes consist in the fact that lone pairs of the lead atoms are smeared out by the near-octahedral symmetry of the lead sites in the high temperature phase, at low temperatures they localize into the traditional lobes.

It is possible to estimate the stereochemical activity of the lone pair from the degree of its delocalization around the atom skeleton. The less the lone pair is delocalized around the atom skeleton, i.e. the higher is the degree of its orbital localized nature in the valency shell, the higher asymmetry it produces in the *p*-element coordination sphere which could develop into complete division into two coordination spheres (I and II). The degree of the stereochemical activity can be characterized by the value of the difference ($\Delta E_1$) between the shortest distances M-X from the spheres I and II. The larger is $\Delta E_1$, the higher stereochemical activity is attributed to the lone pair. Another way of estimation of the lone pair stereochemical activity could be concerned with the value of the difference ($\Delta E_2$) between the shortest distance M-X in the coordination polyhedron of the compound under study and the known shortest distance in this element's compounds, when the lone pair has the highest stereochemical activity. The value of $\Delta E_2$ reflects the degree of delocalization of the lone pair around the atom skeleton as compared to the minimal one. The higher is the $\Delta E_2$ value, the lower is the stereochemical activity of the lone pair in these *p*-element compounds. To estimate the degree of the lone pair stereochemical activity in the compounds with ions $Bi^{3+}$ and $Pb^{2+}$, we found that the minimal distances $d(Bi^{3+}$-O) and $d(Pb^{2+}$-O) in the octahedra are equal to 2.056 Å (in $Bi_8(CrO_4)O_{11}$ [73] and α-$Bi_2B_8O_{15}$ [74]) and 2.400 Å (in $Pb_2V_5O_{12}$ [75]), respectively. Note that the minimal distances $d(Bi^{3+}$-O) are approximately equal to those $d(Bi^{5+}$-O) (2.038 Å in $Sr_4BiO_6(OH)$ [76] and 2.059 Å in $KBiO_3$ [77]) in octahedra of five-valent bismuth which does not have a lone pair and, therefore, its octahedra are not distorted. Search of these distances was performed among the crystal structures of oxides of $Bi^{3+}$ and $Pb^{2+}$ from the Database ICSD; they were the most accurately determined by means of X-ray single crystal diffraction.

We have analyzed the change of $Bi^{3+}$ coordination surrounding at the temperature decrease (see Table 2, Fig. 1(a-d). In the cubic γ-phase $BiFeO_3$ ($Pm\bar{3}m$) (Palai et al. [18]) at 925º C the lone pair is stereochemically inert ($\Delta E_1 = 0$, $\Delta E_2 = 0.766$ Å) and the $Bi^{3+}$ coordination polyhedron comprises a regular cuboctahedron with 12 long Bi-O bonds (d(Bi-O) = 2.82 Å). One should mention that the larger $\Delta E_2$ value in the cubic γ-phase results not only from the lone pair delocalization – an additional $\Delta E_2$ is determined by the sizes of cavities occupied by bismuth ions in the three-dimensional framework of $FeO_6$ octahedra of the perovskite structure. However, Arnold et al. [32] demonstrate that, in contrast with previous reports [18, 19], both the γ- and β-phases exhibit an orthorhombic symmetry (space group *Pbnm*). In the orthorhombic γ-phase at 945º C [32], four nearest oxygen ions (d(Bi-O) = 2.56-2.58 Å) forming the tetrahedron $BiO_4$ flattened along the twofold axis can be singled out in the $Bi^{3+}$ coordination surrounding. In the latter case the lone pair is also virtually stereochemically inert ($\Delta E_1 = 0.02$, $\Delta E_2 = 0.50$ Å) and the $Bi^{3+}$ coordination can not be considered as one-sided.

The transition into the β-phase (at ~930° C) is accompanied with a dramatic increase of the degree of the lone pair stereochemical activity ($\Delta E_1 = 0.20$, $\Delta E_2 = 0.41$ Å at 900° C), whose result is clearly expressed in the presence of a separate group $BiO_3E$ (E – lone pair of electrons) in the в $Bi^{3+}$ coordination. This group can be considered as a distorted tetrahedron with the lone pair in one of its vertices. However, during further temperature decrease down to 830° C ($\Delta E_1$ = 0.16, $\Delta E_2$ = 0.39 Å), one observes, instead of a regular increase of the $\Delta E_1$ value, its decrease indicating to some reduction of the lone pair stereochemical activity, while another parameter ($\Delta E_2$) demonstrates a normal behavior indicating to the activity growth. Prior to discussing the reason of the β-phase instability, one should perform an additional correction of the crystal structure of BiFeO$_3$ in the temperature range under examination. One more dramatic increase of the lone pair stereochemical activity ($\Delta E_1$ = 0.24 Å, $\Delta E_2$ = 0.27 Å) occurs at the transition into the rhombohedral α-phase (space group R3c) at T = 650º C.

**Table 2** Structural parameters, interatomic distances and the degree of the stereochemical activity ($\Delta E_1$ and $\Delta E_2$) of the lone pair in BiFeO$_3$, BiVO$_4$ and PbTiO$_3$.

| Compound, temperature, pressure | Space group, lattice parameters | d((Bi,Pb)-O) (Å) sphere I | d(Bi-O) (Å) sphere II | $\Delta E_1$ (Å) $\Delta E_2$ (Å) | d(Fe-O) (Å) |
|---|---|---|---|---|---|
| α-BiFeO$_3$ [21] 25º C | R3c (rhombohedral) a = 5.579 Å, c = 13.869 Å, γ = 120º | 2.267×3 | 2.534×3 3.208×3 | 0.267 0.211 | 1.953×3 2.105×3 |
| α-BiFeO$_3$ [21] 350º C | R3c a= 5.598 Å, c= 13.935 Å, γ=120º | 2.291×3 | 2.547×3 3.206×3 | 0.256 0.235 | 1.956×3 2.105×3 |
| Bi$_{0.93}$La$_{0.07}$FeO$_3$ [24] | R3c a = 5.573 Å, c = 13.803 Å, γ = 120º | 2.303×3 | 2.508×3 3.198×3 | 0.205 0.247 | 1.962×3 2.089×3 |
| α-BiFeO$_3$ [21] 650º C | R3c a = 5.619 Å, c = 13.982 Å, γ = 120º | 2.322×3 | 2.560×3 3.199×3 | 0.238 0.266 | 1.965×3 2.113×3 |
| β-BiFeO$_3$ [19] 830º C | Pbnm (orthorhombic) a = 5.613 Å, b = 5.647 Å, c = 7.971 Å | 2.449×1 2.482×2 | 2.606×1 2.759×2 2.805×2 | 0.157 0.393 | 2.025×2 2.031×2 2.037×2 |
| β-BiFeO$_3$ [32] 900º C | Pbnm a = 5.630 Å, b = 5.654 Å, c = 7.986 Å | 2.461×2 2.470×1 | 2.659×1 2.774×2 2.845×2 | 0.198 0.405 | 2.031×2 2.034×2 2.037×2 |
| γ-BiFeO$_3$ [32] 945º C | Pbnm a = 5.631 Å, b =5.654 Å, c = 7.989 Å | 2.559×3 | 2.577×1 2.773×2 2.813×2 | 0.018 0.503 | 2.004×2 2.023×2 2.035×2 |
| γ-BiFeO$_3$ [18] 925° C | $Pm\bar{3}m$ (cubic) a = 3.992 Å | 2.822×12 | | 0 0.766 | 1.996×6 |
| BiVO$_4$ [78] -268.5º C | I2/b (monoclinic) a = 5.215 Å, b = 5.084 Å, c = 11.706 Å | 2.314×2 2.349×2 | 2.533×2 2.676×2 | 0.219 0.258 | 1.739×2 1.747×2 |
| BiVO$_4$ [79] 1600 MPa | I4$_1$/a (tetragonal) a = 5.105 Å, c = 11.577 Å | 2.398×2 2.473×2 | 2.398×2 2.473×2 | 0 0.342 | 1.729×4 |
| BiVO$_4$ [78] 293º C | I4$_1$/a a = 5.147 Å, c = 11.722 Å | 2.450×2 2.494×2 | 2.450×2 2.494×2 | 0 0.394 | 1.727×4 |
| PbTiO$_3$ [80] 22º C | P4mm (tetragonal) a = 3.902 Å, c = 4.156 Å | 2.517×4 | 2.798×4 | 0.281 0.117 | 1.770×1 1.979×4 2.386×1 |
| PbTiO$_3$ [81] 2170 MPa | P4mm a = 3.902 Å, c = 4.033 Å | 2.558×4 | 2.780×4 | 0.222 0.158 | 1.766×1 1.970×4 2.266×1 |
| PbTiO$_3$ [80] 427º C | P4mm a = 3.940 Å, c = 4.063 Å | 2.594×4 | 2.805×4 | 0.211 0.194 | 1.808×1 1.985×4 2.255×1 |
| PbTiO$_3$ [81] 2000 MPa, 350º C | P4mm a = 3.930 Å, c = 3.960 Å | 2.632×4 | 2.791×4 | 0.158 0.232 | 1.921×1 1.965×4 2.039×1 |
| PbTiO$_3$ [82] 527º C | $Pm\bar{3}m$ (cubic) a = 3.969 Å | 2.807×12 | | 0 0.407 | 1.985×6 |

To sum up, the analysis showed that the temperature decrease from 945º down to 25º C resulted in the increase of the degree of the stereochemical activity of Bi$^{3+}$ cations lone pair. It was reflected in the increase (from 0.02(*0*) up to 0.27 Å) of the $\Delta E_1$ difference between the shortest distances Bi-O from the spheres I and II due to the decrease (from 0.50(*0.77*) down to 0.21 Å)) of the delocalization degree ($\Delta E_2$) of the lone pair around the Bi skeleton (Table 2, Fig. 1a-d). The italic font is used to show the data for the cubic γ-phase. In Table 2 we present additional data on the changes in the degree of the stereochemical activity of the lone pair under effect of temperature and pressure in BiVO$_4$ [78, 79] and PbTiO$_3$ [80-82] confirming the typical character of the lone pair behavior in BiFeO$_3$.

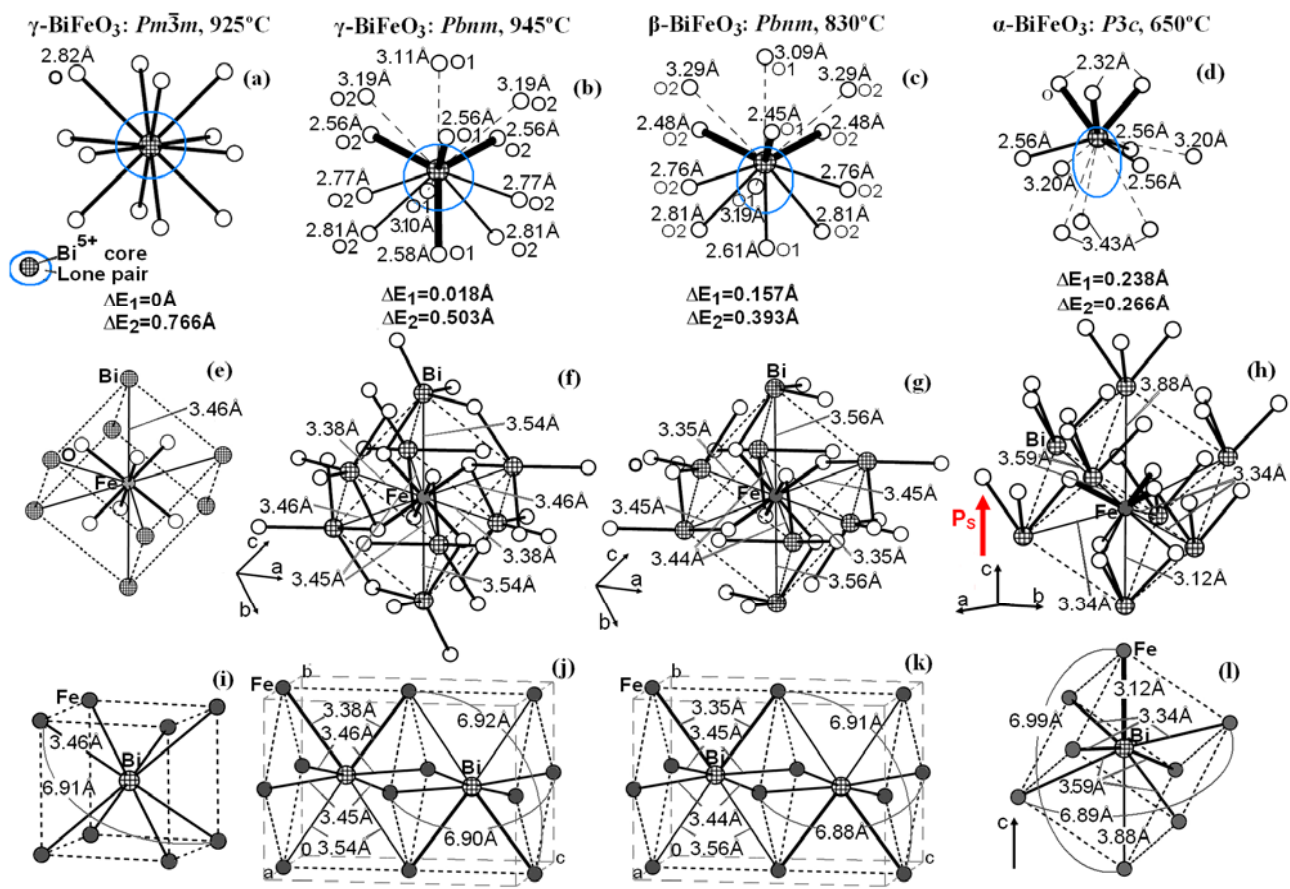

**Fig. 1** Structural effects in BiFeO$_3$ related to reducing the degree of stereochemical activity of the Bi$^{3+}$ lone pair at temperature increase: (a-d) reduction of the difference ($\Delta E_1$) between the shortest distances Bi-O from sphere I and sphere II in the Bi coordination polyhedron; (e-h) breaking of polar orientation of BiO$_3$E groups and electrical polarization of FeO$_6$ octahedra; (i-k) displacement of Bi$^{3+}$ ions relatively to the Fe$^{3+}$ ions sublattice. In this and other figures the thick and thin lines refer to short and long Bi-O, Fe-O bonds and Bi-Fe distances, respectively.

The electronic density of the lone pair always produces the effect of local electrical polarization, since the coordination of cations having the lone pair is one-sided. Orientation of the density of lone pairs in a specific direction might results in electrical polarization of the substance as a whole. However, in most of the compounds the polyhedra with lone pairs are linked to each other through the symmetry center. Besides, the non-centrosymmetry is only a necessary condition for the presence of the polarization, not a sufficient one. There exist compounds whose paraelectric phase is not centrosymmetric. There are cases where the crystal structure is still centrosymmetric but some exotic spin configuration can give rise to the lost of inversion center and emerging of polarization.

It is hard to predict if there will be polarization and in what direction in the compounds containing the lone pair. For example, in BiVO$_4$ (Table 2), where d(Bi-O) = 2.314 Å and $\Delta E_2$ = 0.219 Å, the electrical ordering is not attained even at the temperature down to 4.5 K. One can just speculate that emerging of polarization is preferable along the three-fold axis for the compounds of Bi$^{3+}$ and Sb$^{3+}$ and along the four-fold axis for the Pb$^{2+}$ compounds, since at high activity of the lone pairs they are characterized by formation of groups Bi(Sb)X$_3$E and PbX$_4$E. We have analyzed stoichiometric compounds of Bi$^{3+}$ in the Database ICSD with taking into consideration just one parameter – the presence or absence of the center of symmetry depending on the degree of stereochmical activity of the lone pair estimated on the lengths of the shortest Bi-O distances. It was found out that at d(Bi-O) = 2.056 - 2.155 Å ($\Delta E_2$ = 0 - 0.099 Å) Å only 63 from 225 compounds (28%) and at d(Bi-O) = 2.156 - 2.399 Å ($\Delta E_2$ = 0.100 - 0.343 Å) – 78 from 296 compounds (26.4%) do not have the symmetry center. At the decrease of the stereochemical activity of the lone pair until the stereochemically inert state, confirmed by the increase of the shortest distances Bi-O from 2.400($\Delta E_2$=0.344) Å up to 2.800($\Delta E_2$=0.744) Å, the number of non-centrosymmetric compounds reduces dramatically (4 from 140 (2.9%)).

So the presence of the stereochemically active lone pair guarantees just a local polar distortion of the coordination polyhedron of the cation having the lone pair and, as a secondary effect, distortions of the bonded adjacent polyhedra of other elements. However, there is no any guarantee that there would occur polarization of the structure as a whole. Nevertheless, the analysis we performed demonstrates that increase of the stereochemical activity of the lone pair facilitates the emerging of one of the main characteristics of polarization – noncentrosymmetry of compounds.

## 4 Relation of the paraelectric-ferroelectric phase transition with increase of the lone pair stereochemical activity

Let us consider the structural effects accompanying the electrical polarization in $BiFeO_3$. In the perovskite-type structure of the cubic γ-phase of $BiFeO_3$ the sublattice of $Fe^{3+}$ and $Bi^{3+}$ ions can be represented as a cubic sublattice of the $Fe^{3+}$ ions with the $Bi^{3+}$ ions in the cells centers (Fig. 1i). The $Bi^{3+}$ and $Fe^{3+}$ cations alternate along diagonals (d(Fe-Fe) = 6.914 Å) of the cubic sublattice of $Fe^{3+}$ at equal distances ((d(Fe-Bi) = 3.457 Å)) from each other (Fig. 1e and i, 2a).

In the rhombic γ- and β-phases (Fig. 1j) the diagonal lengths are unequal – two of them (d(Fe-Fe) = 6.922(6.912 ) Å at 945º(830º) C) are longer than others (d(Fe-Fe) = 6.904(6.884) Å at 945º(830º) C). The $Bi^{3+}$ ions slightly deviate from the bond line Fe-Fe. The Fe-Bi-Fe angles are equal to 177.2(175.6)º and 178.4(177.5)º at 945º(830º) C for the short and long diagonal, respectively. Besides, the $Bi^{3+}$ ions displace along the diagonals towards each other, thus approaching one $Fe^{3+}$ ion and moving away from another one (Fig. 2b). The difference between short and long distances Fe-Bi along the short diagonal is insignificant (0.004-0.009 Å), but it attains 0.215 Å for the long one (Fig. 1f, g and j, 2b; Table 2).

During the transition to the rhombohedral α-phase (Fig. 1k and 2c), one of two long diagonals increases abruptly. At 650º C its length (d(Fe-Fe) = 6.991 Å) is by ~0.1 Å larger than that of three other diagonals. The remarkable feature consists in the fact that the direction of this unusually long diagonal coincides with the electrical polarization direction. Let us consider the processes which could occur along the above direction (Fig. 2c).

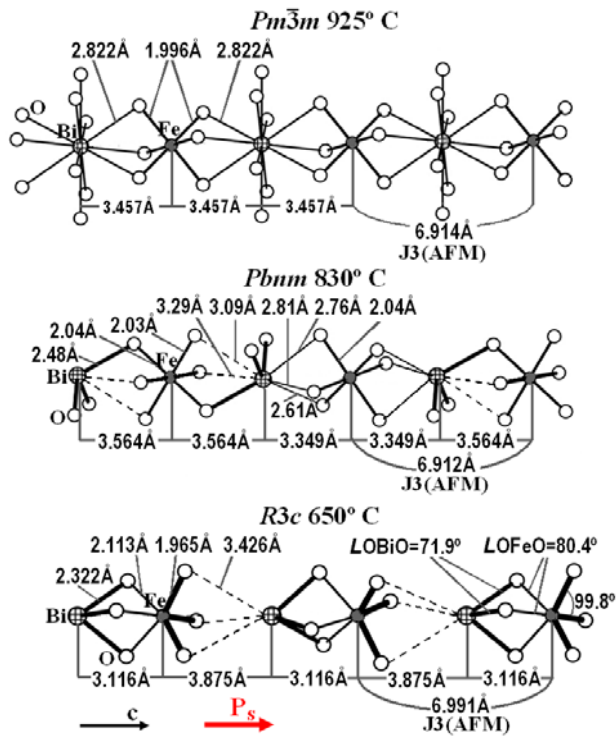

**Fig. 2** Schematic representation of the thermal evolution of the crystal structure of $BiFeO_3$ along the direction of the electrical polarization.

In the β-phase the lone pairs are already stereochemically active (see the section 3.1. above), but oriented in opposite directions at an angle to the long diagonal (Fig. 1g and 2b). With the increase of the degree of the stereochemical activity of lone pairs their influence on the structure also increases. Regarding the long diagonals, the above influence is expressed in the increase of the difference between long and short distances Fe-Bi. The maximum value of this difference (0.215 Å) is reached at 830° C due to decrease (down to 0.393 Å) of the degree of delocalization ($\Delta E_2$) of the lone pair around the $Bi^{3+}$ skeleton. At such a distortion the rhombic phase becomes unstable and undergoes the transition into the rhombohedral α-phase through reorientation of lone pairs in the same direction in parallel to one of two long diagonals which becomes the *c* axis in the α-phase (Fig. 1h and 2c). The latter produces the effect of electrical polarization along the axis *c* of not only the Bi-O sublattice, but also of that of $Fe^{3+}$-O.

Let us examine the structural changes accompanying the reorientation of the lone pairs in the same direction, for the sake of simplicity, by comparing the crystal structures of the α-phase at T = 650°(25°) C and undistorted cubic γ-phase $BiFeO_3$ T = 925° C (Fig. 2a and c). To get positions along the diagonal, the lone pairs act in two ways: first, they displace (by 0.380(0.413) Å at T=650°(25°) C) the $Bi^{3+}$ ions from the middle of the diagonal Fe-Fe in the same direction and, second, make this diagonal longer. Thus, the α-phase undergoes a structural dimerization along the axis *c*. The distances Bi-Fe in dimers from the $BiO_3$ and $FeO_6$ of the octahedron united by common faces are less by ~0.8 Å than between the dimers. The important result of the dimer formation is the polar distortion of $FeO_6$ octahedra that occurs along the same diagonal. Shortening of three Fe-O bonds located at the side of the lone pair can be related to the fact that they become of the end-type due to dimerization and, possibly, due to the pressure of the lone pair, which is corroborated by the increase (by more than 10º) of the angles between these bonds, as compared to the ideal value. Elongation of three opposite bonds Fe-O was the result of their acquired bridging character between the highly charged $Fe^{3+}$ and $Bi^{3+}$ ions during dimerization. Finally, the $Fe^{3+}$ ions appear displaced (by 0.243(0.255) Å at T=650°(25° C) along the axis *c* from the octahedron center to one of its faces.

To conclude on the above, we showed that the reason of the electrical polarization of $BiFeO_3$ consisted in the increase of the degree of the lone pairs' stereochemical activity resulting in their reorientation in one direction and, as the secondary effect, the polarization of $FeO_6$ octahedra.

## 5. Relation of the insulator-metal transition with the lone pair transformation from the stereochemically active to stereochemically inert state

The ß-γ phase transition and the conjugated insulator-metal transition in $BiFeO_3$ are still under intensive investigation. In the recent paper Arnold et al. [32] demonstrate that the insulator-metal phase transition (~930º C) in $BiFeO_3$ takes place without increase of the symmetry within the same orthorhombic space group *Pbnm* and is accompanied by a subtle decrease in the unit cell dimensions and an increase of the Fe-O-Fe bond angle, consistent with an insulator-metal transition. Palai et al. [83] further corroborate the above data by conducting the Raman scattering investigation of ferrite epitaxial thin films, thus disproving the transition to the cubic phase they had established earlier in [18]. Arnold et al. [32] believe that to explain the mechanism of the insulator-metal transition in the absence of space group/symmetry change one can use two alternative models. One of the models [36, 84] is based on the emerging of the high-spin–low-spin transition of $Fe^{3+}$ ions under high pressures. In another model developed for nickelates [85] it is assumed that the insulator-metal transition is due to a closing of the charge-transfer gap between the oxygen *p* orbitals and the iron *d* orbitals. The authors use the similarity of the structural changes accompanying the insulator-metal transition in $BiFeO_3$ and $RNiO_3$ (R = Pr and Nd) as a main proof of the above conclusion.

In our opinion, the disagreement on the symmetry of the paraelectric γ-phase is not concerned with the experimental errors or data interpretation. The studies of crystal chemistry of $BiFeO_3$ and incomplete-valency *p*-elements performed earlier [70, 71, 86] enabled us to conclude that structural unambiguousness of phase transition and the presence of phase transitions, both with and without change of symmetry, is a common peculiarity of the compounds having a lone pair of electrons. This electronic formation creates "non-rigid" sections in the structure making it unstable, since it can easily change its shape and position under the effects of temperature, pressure, at introduction of vacancies or the ion substitution.

We put forward an alternative hypothesis stating that the insulator-metal transition results from the transition of the lone pair from the stereochemically active to the stereochemically inert state at the temperature increase. The comparison of the orthorhombic ß- and γ-phases BiFeO$_3$ at 900º and 945º C (Table 2, Fig. 1b and c) demonstrates that the $Bi^{3+}$ coordination surrounding underwent not only quantitative (bond length change), but also qualitative changes as a result of the dramatic decrease (ΔE$_1$ from 0.20 down to 0.02 Å) of the stereochemical activity of the lone pair down to virtually inert state at 945º C. The $Bi^{3+}$ coordination surrounding acquired, instead of the one-sided "umbrella"-type, the form of a distorted tetrahedron due to elongation (by 0.089 – 0.098 Å) of three Bi-O bonds in the sphere I and shortening (by 0.082 Å) of one of the Bi-O bonds (along the axis *b*) in the sphere II. The latter resulted in shortening by just 0.003, 0.008 and 0.030 Å of lengths of three pairs of Fe-O bonds in FeO$_6$ octahedra and increase by 3º and 5º of the Fe-O-Fe bond angles.

The comparison of the orthorhombic ß-phase at 900º C with the cubic γ-phase at 925º C (´Table 2, Fig. 1a and c) shows that the transformation of the $Bi^{3+}$ coordination surrounding into a cuboctahedron, where its lone pair completely inert, results in a dramatic increase (by 0.352-0.361 Å) of the lengths of three short bonds Bi-O, somewhat smaller increase (by 0.048-0.163 Å) of the next in length three bonds, insignificant shortening (by 0.023 Å) of two long bonds and the coordination increase up to 12 due to nearing four additional oxygen ions. Such substantial changes in the bismuth coordination result in decrease of the average length Fe-O in the octahedron FeO$_6$ by just 0.038 (1.87%) and, at the same time, in significant increase (by 21º and 23º) of the Fe-O-Fe bond angles.

One should mention that the same tetrahedron BiO$_4$ flattened along the twofold axis, just like in the orthorhombic γ-phase of BiFeO$_3$, is formed by the nearest $Bi^{3+}$ ions in BiVO$_4$ [78, 79] (Table 2) at a temperature 293º C and under pressure 1600 MPa. At the same time, PbTiO$_3$ [80-82] (Table 2) undergoes, under the temperature or pressure effect, the transition from the tetragonal to cubic symmetry with the space group $Pm\bar{3}m$ identical to that in the γ-phase of BiFeO$_3$. As a result, the one-dimensional coordination of PbO$_4$E as a tetragonal pyramid, whose one vertex is occupied by the stereochemically active lone pair, transforms into a regular cuboctahedron with the stereochemically inert lone pair. Unfortunately, we could not find the data on the BiVO$_4$ and PbTiO$_3$ electroconductivity at high temperature and pressure.

On the other hand, one can present numerous examples in which the phase transitions insulator-metal or semiconductor-metal occurring under effect of temperature or pressure are accompanied by the change of the electron lone pair behavior. We previously demonstrated [71, 86] on the examples of homology series semiconductor–metal consisting of non-transitional elements and double compounds of the type MeX (Me – metal, X – elements of IV, V and VI groups of the Periodic Table) the existence of correlation between the increase of metallic properties and the decrease of the stereochemical activity of lone pairs of *s*-electrons. In the groups of *p*-elements at moving downward on the Table or under the pressure effect, there occurs, simultaneously with the transition from semiconductor to metal, the transition of *s*-electron pairs from the binding state to the free one with a low degree of the stereochemical activity or to the inert state. For example, under atmospheric pressure α-Sn [87] (Fig. 3a) crystallizes into a cubic system and has the diamond structure, in which each atom is surrounded by four adjacent ones in the regular tetrahedron vertices. During the pressure increase α-Sn transforms into a new tetragonal modification with the structure of the β-Sn-type [88] (Fig. 3b), in which each atom is surrounded by six adjacent ones by the vertices of a slightly distorted octahedron. The distances Sn-Sn in tetrahedra of the α-modification are less by 0.21 Å than the shortest distance in the β-modification octahedra. The cubic modification of α-Sn, in which the pair of *s*-electrons participates in the covalent binding, has semiconductor properties, while the tetragonal modification, in which the pair of *s*-electrons concentrates mainly around the atom skeleton and has a low degree of the stereochemical activity, can be considered as metal. In the semiconductor compound NaSb [89] (structural type LiAs) (Fig. 3c), the Sb coordination polyhedron comprises a distorted eight-capped structure in which two caps are occupied by antimony atoms (two short distances Sb-Sb 2.85 Å) and six remaining caps – by sodium atoms (six long distances Sb-Na 3.13-3.51 Å) that indicates to high stereochemical activity of the antimony lone pair in this compound. During the transition from NaSb to NaBi [90] (structural type CuAu) (Fig. 3d) one can observe a complete restructuring accompanied by the transition from semiconductor properties to metallic ones. Instead of the chain-like grouping of antimony atoms in NaSb, in the compound NaBi there occurs a uniform distribution of Na and Bi atoms on the motif of a tetragonally distorted (*c/a* 0.98) high-density cubic packing characteristic for typically metallic phases. The bismuth lone pair in NaBi is stereochemiclly inert, since the bismuth polyhedron comprises a virtually regular cubic octahedron with four long bonds Bi-Bi 3.46 Å and eight bonds Bi-Na of almost the same size (3.43 Å).

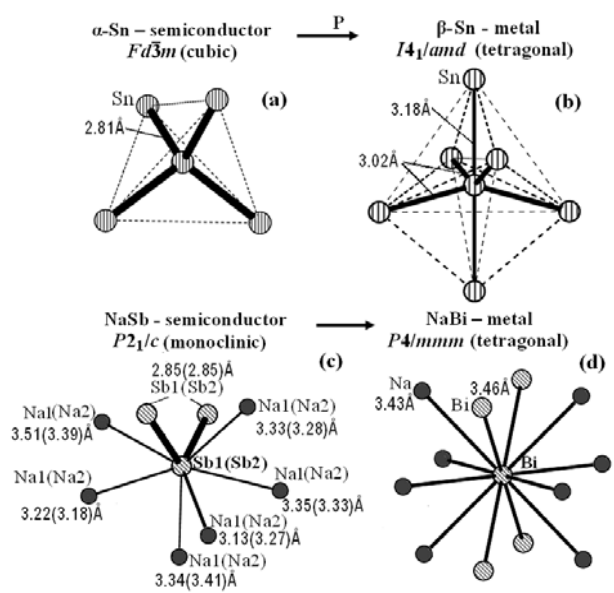

**Fig. 3** Evolution of the coordination surrounding of *p*-elements during the semiconductor-metal transition: (a, b) α-Sn - β-Sn; (c-d) NaSb - NaBi.

The relation of the insulator-metal transition with the lone pair transformation from the stereochemically active to stereochemically inert state was also shown by Lefebvre et al. [68] on the example of antimony chalcogenides and by Waghmare et al. [41] – on the example of chalcogenides of divalent Ge, Sn, and Pb.

To sum up, the analysis of the crystal structure of $BiFeO_3$ under the temperature effect we performed showed that the phase transitions were accompanied by the decrease of the stereochemical activity of the lone pair until its transition into the inert state. Bismuth ferrite is an insulator and ferroelectric at high stereochemical activity of the lone pair, an insulator and paraelectric at the lone pair stereochemical activity decrease, and a metal and paraelectric at the transition into the inert state.

## 6 Relation of the transition magnetic order-magnetic disorder with the changes of the crystal structure at the temperature decrease

The change of the magnetic coupling parameters in $BiFeO_3$ can result from the displacement of intermediate ions in the local space between the iron ions during the change of the behavior of the lone pair of electrons under the temperature effect.

To determine the relation of the above phenomena, we calculated the sign and strength of magnetic couplings in $BiFeO_3$ on the basis of the structural data [18, 19, 21, 24, 32] at temperatures from 25º up to 945º C (Table 3, Fig. 4). To simplify the description, we will examine the orthorhombic γ-phase of $BiFeO_3$ [32] along with the orthorhombic β-phase, since at temperatures 830º, 900º and 945º C the sign of respective couplings does not change while their strength, according to our data, vary only slightly.

The crystal sublattice of the magnetic ions $Fe^{3+}$ comprises a cubic lattice in the cubic γ-$BiFeO_3$ ($a = 3.992$ Å, α= 90º at a temperature 925±5° C) and a slightly distorted cubic lattice in the orthorhombic γ- and β-$BiFeO_3$ ($a = b = 3.981-3.990$ Å, $c = 3.986-3.994$ Å, γ = 89.66 – 89.77º in the temperature range 830 – 945° C) and in the rhombohedral α-$BiFeO_3$ ($a = 3.965 – 3.994$ Å, α= 90.57 – 90.61º in the temperature range 25 – 650° C).

According to our calculations (Table 3), the temperature decrease produces dramatic changes in two types of magnetic couplings. The latter include the couplings in chains along the tetragonal axes of the cubic sublattice of $Fe^{3+}$ with the second ($J1_2$) and third neighbors ($J1_3$) of the iron ion and the couplings ($J2_1$) along the cube faces diagonals, including those located in the planes perpendicular to the polarization direction.

Both of the above coupling types are capable to create magnetic disordering depending on the type of the magnetic moments orientation.

**Table 3** Parameters of magnetic couplings in BiFeO$_3$ calculated on the basis of structural data.

| | γ-BiFeO$_3$ [18] 925° C $Pm\bar{3}m$ | γ-BiFeO$_3$ [32] 945º C $Pbnm$ | β-BiFeO$_3$ [32] 900º C $Pbnm$ | β-BiFeO$_3$ [19] 830º C $Pbnm$ | α-BiFeO$_3$ [21] 650º C $R3c$ | α-BiFeO$_3$ [21] 350º C $R3c$ | α-BiFeO$_3$ [21] 25º C $R3c$ | Bi$_{0.93}$La$_{0.07}$FeO$_3$ [24] $R3c$ |
|---|---|---|---|---|---|---|---|---|
| d(Fe-Fe) (Å) | 3.992 | 3.990 | 3.989 | 3.981 | 3.994 | 3.980 | 3.965 | 3.956 |
| angle FeOFe | 180° | 162.2° | 157.04° | 157.12° | 156.8° | 155.96° | 155.36° | 155.06 |
| $J1_1$ (Å$^{-1}$) | -0.1757 | -0.1367 | -0.1250 | -0.1258 | -0.1210 | -0.1194 | -0.1191 | -0.1195 |
| d(Fe-Fe) (Å) | 7.983 | 7.980 | 7.978 | 7.962 | 7.988 | 7.960 | 7.929 | 7.911 |
| $J1_2$ (Å$^{-1}$) | -0.0348 | -0.0319 | -0.0307 | -0.0307 | -0.0134 | 0.0040 | 0.0039 | -0.0135 |
| d(Fe-Fe) (Å) | 11.975 | 11.970 | 11.969 | 11.943 | 11.983 | 11.939 | 11.894 | 11.867 |
| $J1_3$ (Å$^{-1}$) | -0.0279 | -0.0228 | -0.0213 | -0.0214 | -0.0108 | 0.0005 | 0.0005 | -0.0200 |
| d(Fe-Fe) (Å) | | 3.994 | 3.994 | 3.986 | | | | |
| angle FeOFe | | 161.58° | 158.94° | 157.87° | | | | |
| $J1'_1$ (Å$^{-1}$) | | -0.1349 | -0.1291 | -0.1272 | | | | |
| d(Fe-Fe) (Å) | | 7.989 | 7.986 | 7.971 | | | | |
| $J1'_2$ (Å$^{-1}$) | | -0.0314 | -0.0310 | -0.0309 | | | | |
| d(Fe-Fe) (Å) | | 11.983 | 11.979 | 11.957 | | | | |
| $J1'_3$ (Å$^{-1}$) | | -0.0224 | -0.0215 | -0.0215 | | | | |
| d(Fe-Fe) (Å) | 5.645 | 5.646 | 5.644 | 5.633 | 5.679 | 5.659 | 5.634 | 5.614 |
| $J2_1$ (Å$^{-1}$) | 0.0005 | 0.0003 | 0.0004 | 0.0001 | 0.0006 | 0.0004 | 0.0001 | 0.0001 |
| d(Fe-Fe) (Å) | 11.290 | 11.292 | 11.289 | 11.267 | 11.357 | 11.318 | 11.268 | 11.229 |
| $J2_2$ (Å$^{-1}$) | -0.0093 | -0.0120 | -0.0099 | -0.0103 | 0.0006 | 0.0002 | -0.0001 | -0.0001 |
| | | -0.0041 | -0.0053 | -0.0060 | | | | |
| d(Fe-Fe) (Å) | | 5.646 | 5.644 | 5.633 | | | | |
| $J2'_1$ | | 0.0004 | 0.0005 | 0.0003 | | | | |
| d(Fe-Fe) (Å) | | 11.292 | 11.289 | 11.267 | | | | |
| $J2'_2$ (Å$^{-1}$) | | -0.0101 | -0.0091 | -0.0091 | | | | |
| | | -0.0064 | -0.0061 | -0.0060 | | | | |
| d(Fe-Fe) (Å) | | 5.631 | 5.630 | 5.613 | 5.619 | 5.598 | 5.579 | 5.573 |
| $J2^a_1$ (Å$^{-1}$) | | 0 | -0.0004 | -0.0002 | 0.0015 | 0.0014 | 0.0011 | 0.0009 |
| d(Fe-Fe) (Å) | | 11.262 | 11.260 | 11.227 | 11.237 | 11.195 | 11.158 | 11.147 |
| $J2^a_2$ (Å$^{-1}$) | | -0.0086 | -0.0082 | -0.0082 | 0.0015 | 0.0012 | 0.0007 | 0.0005 |
| d(Fe-Fe) (Å) | | 5.654 | 5.654 | 5.647 | | | | |
| $J2^b_1$ (Å$^{-1}$) | | -0.0002 | -0.0006 | -0.0007 | | | | |
| d(Fe-Fe) (Å) | | 11.308 | 11.307 | 11.294 | | | | |
| $J2^b_2$ (Å$^{-1}$) | | -0.0090 | -0.0087 | -0.0086 | | | | |
| d(Fe-Fe) (Å) | 6.914 | 6.922 | 6.922 | 6.912 | 6.991 | 6.968 | 6.935 | 6.901 |
| $J3$ ($J_{c/2}$) | -0.0432 | -0.0409 | -0.0411 | -0.0402 | -0.0373 | -0.0378 | -0.0384 | -0.0384 |
| d(Fe-Fe) (Å) | | 6.904 | 6.902 | 6.884 | 6.894 | 6.868 | 6.844 | 6.834 |
| $J3'$ (Å$^{-1}$) | | -0.0396 | -0.0404 | -0.0388 | -0.0230 | -0.0225 | -0.0223 | -0.0250 |
| d(Fe-Fe) (Å) | 8.925 | 8.929 | 8.927 | 8.910 | 8.893 | 8.860 | 8.830 | 8.819 |
| $J4$ (Å$^{-1}$) | -0.0373 | -0.0301 | -0.0283 | -0.0285 | -0.0272 | -0.0271 | -0.0271 | -0.0282 |
| | | | | | -0.0303 | -0.0302 | -0.0304 | -0.0307 |
| d(Fe-Fe) (Å) | | 8.929 | 8.927 | 8.910 | 8.969 | 8.938 | 8.900 | 8.871 |
| $J4'$ (Å$^{-1}$) | | -0.0315 | -0.0294 | -0.0296 | -0.0259 | -0.0257 | -0.0257 | -0.0259 |
| | | | | | -0.0344 | -0.0347 | -0.0347 | -0.0347 |
| d(Fe-Fe) (Å) | 9.777 | 9.774 | 9.771 | 9.749 | 9.767 | 9.731 | 9.695 | 9.677 |
| $J5$ (Å$^{-1}$) | 0.0377 | 0.0113 | 0.0038 | 0.0037 | 0.0137 | 0.0131 | 0.0127 | 0.0125 |
| $J5'$ (Å$^{-1}$) | | 0.0075 | 0.0080 | 0.0081 | 0.0175 | 0.0165 | 0.0158 | 0.0155 |
| d(Fe-Fe) (Å) | | 9.787 | 9.785 | 9.769 | 9.732 | 9.696 | 9.663 | 9.653 |
| $J5''$ (Å$^{-1}$) | | 0.0092 | 0.0050 | 0.0070 | 0.0234 | 0.0229 | 0.0223 | 0.0193 |
| d(Fe-Fe) (Å) | | 9.787 | 9.785 | 9.769 | 9.870 | 9.836 | 9.791 | 9.748 |
| $J5'''$ (Å$^{-1}$) | | 0.0110 | 0.0095 | 0.0092 | 0.0119 | 0.0112 | 0.0107 | 0.0100 |

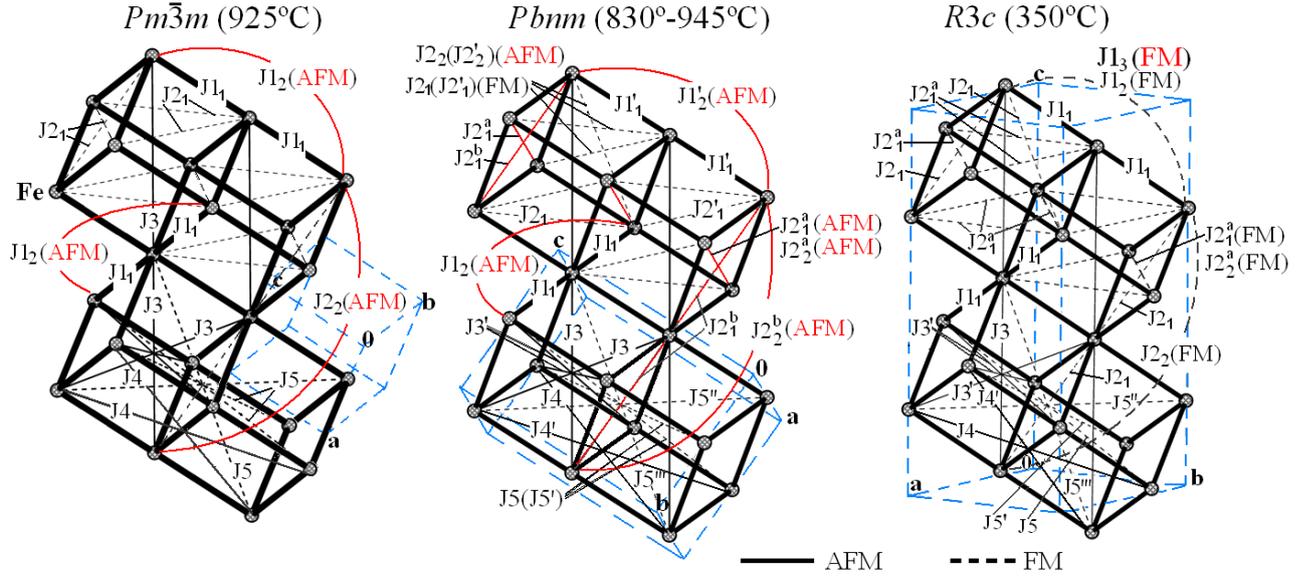

**Fig. 4** The sublattice of $Fe^{3+}$ and the coupling $J_n$ in cubic, orthorhombic and rhombohedral phases of $BiFeO_3$. In this and other figures the thickness of lines shows the strength of $J_n$ coupling. AFM and FM couplings are indicated by solid and dashed lines, respectively.

Let us consider the couplings along the tetragonal axes of the cubic sublattice of the iron ions. $J1_1$ – magnetic couplings of each $Fe^{3+}$ with its six nearest neighbors remain dominating strong AFM-couplings before and after magnetic ordering. Nevertheless, the strength of these couplings ($J1_1^\gamma = 1.29 J1_1^{\gamma(945C)} = 1.39 J1_1^{\beta(830C)} = 1.41 J1_1^{\alpha(650C)} = 1.42 J1_1^{\alpha(25C)}$, $J1_1^\gamma = -0.176$ Å$^{-1}$) decreases dramatically during transition from the cubic to the orthorhombic phase and then more smoothly with the temperature decrease. Substantial contribution of $j_O$ to the AFM-component of $J1_1$ couplings emerges under effect of only bridging oxygen atoms linking the magnetic ions into chains –Fe-O-Fe-O- along the parameters of the Fe sublattice (Fig. 5). In the cubic phase the bridging oxygen ion is located immediately in the middle of the bond line Fe-Fe ($h$(O) = 0), but, along with the temperature decrease, it is more and more displaced from the bond line to the boundary of the local interaction space at distances $h$(O) = 0.312 - 0.405 Å in the β-phase and $h$(O) = 0.409 - 0.432 Å in the α-phase. It is clearly seen in the increase of the $FeO_6$ octahedra inclination (the angle FeOFe is equal to 180° in the cubic γ-phase, decreases from 162.2° down to 157.1° in the β-phase and from 156.8° to 155.4° in the α-phase). Increase of the displacement of the bridge oxygen atoms results in consecutive decrease of the value of AFM $j_O$ contributions from -0.176 Å$^{-1}$ in the γ-phase to -0.137 – -0.126 Å$^{-1}$ in the β-phase and, further, to -0.125 - -0.124 Å$^{-1}$ in the α-phase.

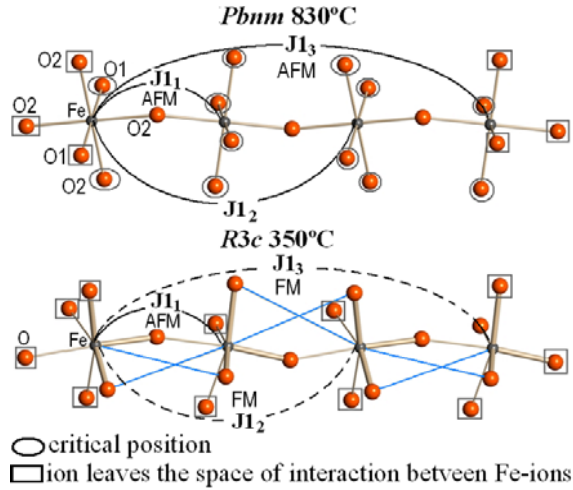

**Fig. 5** The arrangement of intermediate ions in local spaces of $J1_1$ $J1_2$ and $J1_3$ couplings.

During the transition into the β-phase, there is an additional displacement of four oxygen atoms into the local space of $J1_1$ couplings (Fig. 5), however, they do not initiate the emerging of the magnetic coupling ($j_O$ = 0), since they are located near its boundaries ($h(O)$ = 1.988 – 2.025 Å). As a result of polar displacements accompanying the electrical ordering at the transition to the α-phase, only two of them markedly (~0.1 Å) enter inside the local space and initiate small contributions to the coupling FM-component, because they are located not in the central one-third of the local space, but near the magnetic ions ($l'/l$ = 11-52). After summing up the contributions with opposite signs their FM-components only slightly reduce the value of the $J1_1$ coupling.

However, these additional ions have a crucial role in determination of the spin orientation in $J1_2$ and $J1_3$ couplings with the second and third neighbors of the iron ion along the cubic sublattice parameters. The point is, the above ions get into the central one-third ($l'/l$ < 2) of the local space of the $J1_2$ and $J1_3$ couplings and, therefore, can initiate large FM-contributions. This immediately happens to the first oxygen atom as a result of the ferroelectric transition, since it advances to the bond line Fe-Fe at a distance $h(O)$ = 1.927 Å as early as at T = 650° C. It is more difficult to determine the beginning of initiation of coupling by the second oxygen ion. After the ferroelectric transition at T = 650° C it continues to be on the bond line at a distance $h(O)$ = 1.963 Å, which surpasses the critical value ($h_c(O)$ = ~1.95 Å). The increase of the structure distortion in the same direction at further temperature decrease gradually moves this oxygen ion to the interaction line ($h(O)$ is equal to 1.958 Å, 1.957 Å, 1.955 Å and 1.956 Å at temperatures 550°, 450° C, 390° C and 370° C, respectively). One can assume that the second oxygen ion starts to initiate the magnetic ordering only at a temperature between 370° C and 350° C, when the distance $h(O)$ reduces down to 1.954 Å. This distance remains constant down to 200° C and then proceeds to again slowly reduce down to 1.951 Å at the temperature decrease down to 25° C.

The order of the additional oxygen ions entering into the space of interaction with the second and third neighbors is the same as for the first neighbors, however, the number of oxygen atoms doubles and triples and, moreover, there is an addition of one and two iron ions, respectively. In the cubic and orthorhombic paraelectric phases $J1_2(J1'_2)$ and $J1_3(J1'_3)$ the couplings with the second and third neighbors are antiferromagnetic, because they are formed under the effect of the intermediate $Fe^{3+}$ ions and only the bridge oxygen ions located on the Fe-Fe bond line or in its vicinity and making AFM-contributions. The strength of these couplings is significantly weaker than that of $J1_1(J1'_1)$ couplings ($J1_2/J1_1$ = 0.20 and $J1_3/J1_1$ = 0.11 in the γ-phase, $J1_2/J1_1(J1'_2/J1'_1)$ = 0.23 - 0.25 and $J1_3/J1_1(J1'_3/J1'_1)$ = 0.17 in the β-phase). The AFM couplings $J1_2(J1'_2)$ with the second neighbors compete with the nearest AFM couplings $J1_1(J1'_1)$. This competition preserves after the electrical polarization at further temperature decrease down to ~460°, since the spin orientation does not change in couplings with the second neighbors. The reason is in the fact that the sum of the FM contributions from the first "batch" of additional oxygen atoms entering the local sphere of interactions after polar displacements is not sufficient to surpass the AFM contributions from the intermediate $Fe^{3+}$ ions and bridge oxygen atoms. The FM contributions from the first "batch" of additional ions just reduce two-fold ($J1_2/J1_1$ = 0.11 and $J1_3/J1_1$ = 0.09 in the α-phase at T = 650° C) the strength of couplings with the second and third neighbors. The competition between the $J1_1$ and $J1_2$ couplings ceases to exist only after emerging of the FM contributions from the second "batch" of additional oxygen ions which enter the interaction space at further structure distortion in the same direction as a result of the increase of the stereochemical activity of the lone pair with the temperature decrease down to 370° C – 350° C. The sum of the above contributions, along with the FM contributions from the first „batch" of additional ions, finally surpasses the sum of the AFM contributions resulting in the transition of the $J1_2$ ($J1_2/J1_1$ = -0.03 at T = 350° C) couplings from the AFM to the FM state. This very fact appears to be the crystal chemistry reason of the large difference between the temperatures of ferroelectric and antiferromagnetic orderings in $BiFeO_3$. Since all the oxygen atoms in the perovskite-type structure are of the bridge type, entering of additional oxygen ions into the $J1_2$ couplings space correlates in $BiFeO_3$ with the Fe-O-Fe bond angle, and the emerging of the magnetic ordering can be derived from this angle value (Fe-O-Fe≤155.96°). However, after the Bi substitution by La the magnetic ordering does not occur even at an angle of 155.06°.

Simultaneously with the $J1_2$ couplings and due to the same reason, the $J1_3$ couplings with the third neighbors along the cubic sublattice parameters undergo the AFM→FM transition, but their strength reduces 20-fold. As a result, even after attaining the magnetic ordering temperature, there remains a weak competition along the cubic sublattice parameters between the FM $J1_3$ and AFM $J1_1$ couplings which does not disappear with the temperature decrease. This competition must be one of the reasons of the emerging of the non-collinear magnetic structure at temperatures 350°-370° C. It seems to be impossible to establish the configuration of this non-collinear magnetic structure from our data.

According to our calculations of the magnetic couplings parameters in $Bi_{0.93}La_{0.07}FeO_3$ [24] (Table 3), the suppression of this spatially modulated spin structure at Bi substitution by La takes place on two reasons. First, the orientation of magnetic moments of all couplings remains the one they had in the non-substituted $BiFeO_3$ at a temperature 650º C, when the electrical polarization already occurred, while the magnetic ordering did not. Second, simultaneously, one can observe about two-fold increase of the strength of the AFM $J1_3$ couplings with the third neighbor along the tetragonal axes of the cubic sublattice. In this case the application of the magnetic field might cause the transition to the FM state for only the $J1_2$ couplings with the second neighbors, while the $J1_3$ couplings with the third neighbors would remain antiferromagnetic and not competitive with the AFM $J1_1$ couplings.

Let us consider the couplings along the cube faces diagonals which also change the ordering type under the temperature effect. There are no bridge oxygen ions between the iron ions located along the cube faces diagonals. In the cubic γ-phase all the couplings along the cube faces diagonals are equivalent, while in the rhombic β-phase they are divided into four types which we named as $J2_1$, $J2'_1$, $J2^a_1$ (along the axis *a*) and $J2^b_1$ (along the aixs *b*) and in the rhombohedric – into two types $J2_1$ and $J2^a_1$ (along the axes *a* and *b*) (Fig. 4). In the local couplings space of $J2_1$, $J2'_1$, $J2^a_1$ and $J2^b_1$ with the first neighbors there are contained four oxygen atoms in each of γ- and β-phases and after the transition to the α-phase there appears an extra oxygen ion for any system, while for the second neighbors ($J2_2$, $J2'_2$, $J2^a_2$ и $J2^b_2$) the number of oxygen atoms doubles and one extra iron ion is also added. All the above couplings are weak and unstable (Table 3), because the sum of contributions of these ions into the AFM and FM interaction components are approximately equal (critical position 'd'). The strength of the couplings with the second neighbors is just slightly higher than that with the first neighbors, taking into account that with the increasing distance the coupling strength falls with an ever-growing rate. In the γ-phase all the $J2_1$ couplings with the first neighbors are of the FM type, while in the β-phase a half of them ($J2^a_1$ and $J2^b_1$) transform into the AFM type. The latter participate in the competition with the second-neighbor couplings ($J2_2$, $J2'_2$, $J2^a_2$ and $J2^b_2$) which are in both paraelectric phases of the AFM-type due to the AFM contribution from the iron ions. All the AFM couplings along the diagonals of the cubic lattice faces, including those located in the plane perpendicular to the polarization direction ($J2_1$, $J2'_1$ and $J2^a_1$ in the β-phase correspond to $J2^a_1$ in the α-phase) transform into the FM state immediately after the electrical ordering due to the FM contributions from the additional oxygen ions entering the local interaction space and cease to compete with each other. However, at the temperature decrease down to 25° C the $J2_2$ couplings in the α-phase transform into the AFM type and become capable to participate in the formation of the spatially modulated spin structure. One should mention that the transition of the $J2_2$ couplings into the AFM state occurs not in all samples of $BiFeO_3$. For example, according to our calculations, the $J2_2$ couplings preserve their FM at room temperature in the $BiFeO_3$ sample whose structure is presented in [23].

Other couplings do not change the spin ordering under the temperature effect (Table 3). The couplings along the cube diagonals ($J3$ and $J3'$) and diagonals of the faces of the parallelepiped ($J4$ and $J4'$) formed by two cubes are of the AFM type, while the couplings along the diagonals of the above parallelepiped ($J5$ and $J5'$) are of the FM type. One should mention that the couplings along the polarization direction with the first ($J3_1$) and the second ($J3_2$) neighbors d(Fe-Fe = $c/2$ и $J3_2 = c$ in the α-phase) do not compete with each other, since they are of the AFM- and FM-types, respectively.

To sum up, in two paraelectric phases – cubic and orthorhombic – the AFM $J1_2(J1'_2)$ and $J2_2(J2'_2, J2^a_2$ and $J2^b_2)$ couplings with the second neighbors compete with the nearest couplings along the tetragonal axes of the cubic sublattice (AFM $J1_2(J1'_2)$ - AFM $J1_1(J1'_1)$) and along the cube edges diagonals (AFM $J2_2(J2'_2)$ - FM $J2_1(J2'_1)$, AFM $J2^a_2(J2^b_2)$ - AFM $J2^a_1(J2^b_1)$). Besides, in the cubic phase the AFM $J1_2$ and $J2_2$ couplings compete with other couplings in the following triangles: AFM $J1_2$ - FM $J2_1$- FM $J2_1$, AFM $J1_2$ - AFM $J1_1$ - AFM $J4$, AFM $J1_2$ - FM $J2_1$ - FM $J5$, AFM $J2_2$ - AFM $J3$ - AFM $J3$. In the orthorhombic phase these AFM $J1_2(J1'_2)$ and $J2_2$ couplings and the transformed ($J2^a_1$ and $J2^b_1$) from the FM into the AFM part of $J2_1$ couplings also include into the competition all the couplings, except the FM $J5(J5'$, $J5''$, $J5''')$, in the following triangles: AFM $J1_2(J1'_2)$ - FM $J2_1$ - FM $J2_1$, AFM $J1'_2$ - AFM $J1_1$ - AFM $J4(J4')$, AFM $J2_2$ - AFM $J3$ - AFM $J3'$, AFM $J2^a_1(J2^b_1)$ - AFM $J1_1$ - AFM $J1_1$, AFM $J2^a_1(J2^b_1)$ - AFM $J1'_1$ - AFM $J3(J3')$, AFM $J2^a_1$ - FM $J2_1$ - FM $J2_1$.

During the transition to the ferroelectric rhombohedral α-phase the nearest-neighbor $J2^a_1$ and the next-nearest-neighbor $J2^a_2$ couplings along the cube faces diagonals become of the FM-type and, as a result, the number of competing couplings decreases, however, the competition preserves and is supported by the AFM $J1_2$ couplings – along the tetragonal axes of the cubic sublattice (AFM $J1_2$ - AFM $J1_1$) and in the following triangles: AFM $J1_2$ - FM $J2_1$ - FM $J2^a$, AFM $J1_2$ - AFM $J1_1$- AFM $J4(4')$, AFM $J1_2$ - FM $J2^a_1$ - FM $J5(J5')$ and AFM $J1_2$ - FM $J2_1$ - FM $J5''(J5''')$. Only at the decrease of the temperature down to 370° C - 350° C there occurs the transition of the AFM ordering of the $J1_2$ couplings into the FM ones and, as a result, the magnetic competition in $BiFeO_3$ ceases to exist. Nevertheless, the emerged AFM→FM transition of the $J1_3$ couplings with the third neighbors preserves small competition (AFM $J1_1$ - FM $J1_2$ - FM $J1_3$) along the cubic lattice parameters. With the temperature decrease down to 25° C the competition slightly increases due to the transition into the AFM of a part of the $J2_2$ couplings with the second neighbors along the cube faces diagonals and reappearance of the competition between the AFM $J2_2$ - FM $J2_1$ and AFM $J2_2$ - AFM $J3$ - AFM $J3'$. It should be specially emphasized that since the competition is formed by the magnetic neighbors located at different distances and having different coupling strengths, the limit case of the magnetic competition – frustration – is not attained. The competition between these interactions can be the reason of the emerging of a magnetic structure with a period incommensurable with the crystal lattice period. We attempted to eliminate the competition along the cubic lattice parameters between the AFM $J1_1$ and FM $J1_3$ couplings and that along the cube diagonals between the FM $J2_1$ and AFM $J2_2$ couplings by varying coordinates (displacement) of the iron ions and/or oxygen within the frames of the space group $R3c$. However, we did not manage to attain a simultaneous transition FM → AFM for the $J1_3$ coupling and the AFM → FM for the $J2_2$ couplings. At any time the competition was eliminated along the cubic lattice parameters, it appeared along the cube faces diagonals and vice versa. Similar phenomenon was observed at studying the multiferroic $TbMnO_3$ [91].

## 7 Conclusions

We have considered the evolution of the crystal structure of $BiFeO_3$ at the increase of the degree of the stereochemical activity of the lone pair of $Bi^{3+}$ electrons with the temperature decrease from 945° down to 25° C. According to our studies, the metal-insulator and paraelectric-ferroelectric transitions in $BiFeO_3$ result directly from the change of the degree of the stereochemical activity of the lone pair – its consecutive increase with the temperature decrease. The structural ambiguity at the ß-γ phase transitions and the presence of phase transitions both with and without the symmetry change is also related to the lone pair effect. This electronic entity creates "non-rigid" fragments in the structure and makes it unstable, because the lone pair can easily change its shape and location under the effects of temperature and pressure, during introduction of vacancies and ions substitution.

The analysis of crystal structures of $Bi^{3+}$ compounds demonstrated that the presence of the stereochemically active lone pair guaranteed only local distortion of the coordination polyhedron of the cation having such a lone pair as well as the distortion of adjacent polyhedra of other cations. However, there is no guarantee that there will occur the electrical polarization of the system as a whole, since in most of the compounds the polyhedra with lone pairs are linked to each other through the symmetry center. High degree of the lone pair stereochemical activity increases the probability of the emerging ferroelectricity.

The emerging of the magnetic ordering is a secondary effect of the change of the stereochemical activity of the lone pair. We calculated the sign and strength of magnetic couplings in $BiFeO_3$ on the basis of structural data in the temperature range from 25° up to 945° C and found the couplings undergoing the AFM→FM transition under the temperature effect. The change of the parameters of the couplings results from the displacements of intermediate ions in the local space between the iron ions during change of the degree of the stereochemical activity of the lone pair. Based on the obtained data and the analysis of the geometric competition of the magnetic couplings, we determined the temperature (~350°-370° C) of the emerging of the AFM ordering and found the reason of the substantial difference of the temperatures of ferroelectric and magnetic orderings. Besides, it was revealed that, simultaneously with the magnetic ordering, there emerges a new very weak competition, which does not disappear with the temperature decrease. This very competition must induce the emerging of the non-collinear magnetic structure of $BiFeO_3$.

We calculated the parameters of the magnetic couplings in $Bi_{0.93}La_{0.07}FeO_3$ and determined possible reasons for the suppression of this spatially modulated spin structure at Bi substitution by La.

**Acknowledgments** This work is supported by grant 09-I-P18-03 of the Far Eastern Branch of the Russian Academy of Sciences.

E-mail: volkova@ich.dvo.ru